\documentclass{ws-ijmpa}
\def\sb{\mbox{\rule{0pt}{11pt}}}
\def\ga{\gamma}

\begin{document}

\markboth{S. F. Radford, W. W. Repko}
{Describing Recently Discovered Narrow States as Quarkonia Using a Potential Model}

%
\catchline{}{}{}{}{}
%

\title{DESCRIBING RECENTLY DISCOVERED \\NARROW STATES AS QUARKONIA\\
USING A POTENTIAL MODEL\footnote{Presented at DPF2004, August 30, 2004}}

\author{\footnotesize STANLEY F. RADFORD}

\address{Department of Physics, Marietta College\\
Marietta, Ohio 45750, USA}

\author{WAYNE W. REPKO}

\address{Department of Physics and Astronomy, Michigan State University\\
East Lansing, Michigan, 48824, USA}

\maketitle


\begin{abstract}
We examine to what extent several recently discovered narrow resonances can be interpreted as conventional $c\bar{c}$ bound states describable using a potential model. In doing so, we use a semirelativistic approach, which includes both the $v^2/c^2$ and QCD one-loop corrections to the short distance potential and a long range linear potential together with its scalar and vector $v^2/c^2$ spin-dependent terms. 
\end{abstract}

\section{Introduction}  

With the recent experimental results for several expected states ($\eta'_C$ and $h_C$)  in the charmonium spectrum, and the discovery of a state ($X(3872)$), which could be a $^3D_2$ charmonium level, it seems an appropriate time to revisit the potential model interpretation of the $c\bar{c}$ spectrum. For such models, the challenges seem to be:
\begin{itemize}  
\item Are potential models capable of describing the spin splitting in a quantitatively satisfactory way?
\item Including the additional data, is it possible to determine the Lorentz properties of the phenomenological confining potential? 
\item How well are the leptonic and radiative decays predicted?
\item Under what circumstances can the state at 3872 MeV be interpreted as a charmonium level?
\end{itemize}
Here, we will attempt to answer these questions using a potential model which includes the $v^2/c^2$ and all one-loop corrections to the short distance potential supplemented with a linear phenomenological confining potential and its  $v^2/c^2$ corrections. 

\section{Potential Models}

Potential models range from non-relativistic forms such as the Cornell model \cite{cornell}
\begin{equation}\label{corn}
\mathcal{H}_C=\frac{p^2}{2m}+Ar-\frac{4\alpha_S}{3r}\,,
\end{equation}
recently used to identify new charmonium states observable in $B$ decays \cite{elq}, to those including all $v^2/c^2$ spin-dependent corrections, \cite{prs,schnit} $\mathcal{H}=\mathcal{H}_C+V_{HF}+V_{LS}+V_{TEN}$, which can be used to analyze the spectrum and decays of charmonium \cite{barnes}. One-loop QCD corrections can then be added to complete the non-relativistic treatment.\cite{grr}

In our analysis, we utilize a semi-relativistic Hamiltonian of the form
\begin{equation}\label{semi}
\mathcal{H}=2\sqrt{p^2+m^2}+Ar-\frac{4\alpha_S}{3r}F(\mu,r)+V_S+V_L= \mathcal{H}_0+V_S+V_L\,.
\end{equation}
Explicit forms of the short distance potential $V_S$, the long distance potential $V_L$ and $F(\mu,r)$, the one-loop QCD correction to $4\alpha_S/3r$, can be found in Ref.\,\refcite{gjrs}. The $c\bar{c}$ mass spectrum and the corresponding wave functions were obtained using a variational approach. The wave functions were expanded as
\begin{equation}\label{wavefun}
\psi^m_\ell(\vec{r})=\sum_{n=0}^N C_n\left(\frac{r}{R}\right)^n e^{-r/R} Y_\ell^m(\Omega)\,,
\end{equation}
and the $C_n$'s were determined by minimizing $E=\langle\psi\,|\mathcal{H}\,|\psi\rangle/\langle\psi\,|\psi\rangle$.
This procedure results in a linear eigenvalue equation for the $C_n$'s and the energies. The wave functions corresponding to different eigenvalues are orthogonal and the $j^{\rm th}$ eigenvalue $\lambda_j$ is an upper bound on the exact energy $E_j$. For $N=10$, the lowest four eigenvalues are stable to a part in $10^6$.

\section{Results and Conclusions}

The energies and wave functions were obtained by treating $V_S$+$V_L$ as a perturbation to ${\cal H}_0$ and by treating ${\cal H}$ nonperturbatively. We fit the spectrum to ten well established \footnote{$1^1S_0,1^3S_1,1^3P_J,2^1S_0,2^3S_1,1^3D_1,3^3S_1$ and $2^3D_1$.} $c\bar{c}$ states by adjusting the parameters $A,\alpha_S,m,\mu$ and the vector fraction $f_V$ to minimize $\chi^2$. Our nonperturbative fit gives: $A=0.175\,\mathrm{GeV}^2,\alpha_S=0.361,m=1.49\,\mathrm{GeV},\mu=1.07\,\mathrm{GeV}$ and $f_V=0.18$. Interestingly, the perturbative fit yields similar results for $A,\alpha_S$ and $m$, but prefers $\mu=2.32\,\mathrm{GeV}$ and $f_V=0$. The results for the levels are given in Table \ref{spectra} and the predicted $E_1$ transitions widths are given in Table \ref{E1}.

The semi-relativistic model provides a quantitatively good description of the charmonium spectrum. Of the states included in the fit, only the $^3D_1(3770)$ is poorly described. The $E_1$ widths agree reasonably well with experiment\cite{dpf2004}. However, based on the model considered here, the $X(3872)$ cannot be explained solely in terms of a charmonium $^3D_2$ state described by a potential. Spin effects alone can only separate the $^3D_2$ from the $^3D_1$ by 40 MeV or so, which suggests that the inclusion of open channel effects is essential if this identification is to be established. 

\begin{table}[h]
\centering
\tbl{Perturbative and nonperturbative results for the $c\bar{c}$ spectrum.} {\begin{tabular}{|c|l|l|l|c|l|l|l|}
\hline
\sb          & Pert   & Non-pert& Expt            &          & Pert  & Non-pert& Expt             \\
\hline
\sb $\eta_c$  & 2985   & 2981    & $2979.7\pm 1.5$ &$\eta'_c$ & 3599  & 3624    &($3637.7\pm 4.4$)\\
\hline
\sb$J/\psi$   & 3096.9 & 3096.9  &$3096.87\pm 0.04$& $\psi'$  & 3686  & 3686    & $3686.0\pm 0.1$ \\
\hline
\sb$\chi_0$   & 3418.4 & 3415.8  &$3415.1\pm 0.8$  & $\chi'_0$& 3849  & 3872    &                 \\
\hline
\sb$\chi_1$   & 3510.2 & 3510.4  &$3510.51\pm 0.12$& $\chi'_1$& 3946  & 3951    &                 \\
\hline
\sb$\chi_2$   & 3556.5 & 3556.3  &$3556.18\pm 0.17$& $\chi'_2$& 3999  & 3996    &                 \\
\hline
\sb$h_c$      & 3527   & 3524    &$(3526.21\pm 0.25)$& $h'_c$   & 3966  & 3966    &                 \\
\hline
\sb$1^3D_1$   & 3809   & 3790    &$3770\pm 2.5$    & $2^3D_1$ & 4174  & 4157    & $4160\pm 20$    \\
\hline
\sb$1^3D_2$   & 3827   & 3826    &$3872\pm 1.0$    & $2^3D_2$ & 4198  & 4201    &                 \\
\hline
\sb$1^3D_3$   & 3831   & 3845    &                 & $2^3D_3$ & 4209  & 4223    &                 \\
\hline
\sb$1^1D_2$   & 3824   & 3825    &$3836\pm 13.0$   & $2^1D_2$ & 4199  & 4202    &                 \\
\hline
\end{tabular}}
\label{spectra}
\end{table}

\begin{table}[h]
\centering
\tbl{$E_1$ transition widths.} 
{\begin{tabular}{|l|c|c|l|c|c|}
\hline
\sb $\Gamma_{\ga}(E1)$\,(keV)& TH & EX &$\Gamma_{\ga}(E1)$\,(keV)& TH & EX \\ 
\hline
\sb$\chi_0\to\ga J/\psi$    & 169 & $119\pm 17$ &$1^3D_2(3826)\to\ga\chi_1$& 314  &  \\
\hline
\sb$\chi_1\to\ga J/\psi$    & 357 & $288\pm 51$ &$1^3D_2(3826)\to\ga\chi_2$& 76.3 &  \\
\hline
\sb$\chi_2\to\ga J/\psi$    & 468 & $426\pm 48$ &$1^3D_2(3872)\to\ga\chi_1$& 459  &  \\
\hline
\sb$h_c\to\ga\eta_c$        & 670 &             &$1^3D_2(3872)\to\ga\chi_2$& 119  &  \\
\hline
\sb$\psi'\to\ga\chi_0$      & 22  & $24.2\pm 2.5$&$\psi(3770)\to\ga\chi_0$ & 291   & $320\pm 100$  \\
\hline
\sb$\psi'\to\ga\chi_1$      & 33  & $23.6\pm 2.7$ &$\psi(3770)\to\ga\chi_1$& 125  & $280\pm 100$ \\
\hline
\sb$\psi'\to\ga\chi_2$      & 29  & $24.2\pm 2.5$ &$\psi(3770)\to\ga\chi_2$& 5.6   & $\leq 330$ \\
\hline
\sb$\eta'_c\to\ga h_c$      & 22  &               &                      &      &           \\
\hline
\end{tabular}}
\label{E1}
\end{table}

\section*{Acknowledgments}

This research was supported in part by the National Science Foundation under Grant PHY-0244789.


\end{document}